# In Need of Creative Mobile Service Ideas? Forget Adults and Ask Young Children

Ilona Kuzmickaja[1]
i.kuzmickaja@gmail.com
Faculty of Computer Science, Free University of Bozen-Bolzano

Xiaofeng Wang[1]
xiaofeng.wang@unibz.it
Faculty of Computer Science, Free University of Bozen-Bolzano

Daniel Graziotin[1]
**Corresponding Author**
daniel.graziotin@unibz.it
Faculty of Computer Science, Free University of Bozen-Bolzano

Gabriella Dodero
gabriella.dodero@unibz.it
Faculty of Computer Science, Free University of Bozen-Bolzano

Pekka Abrahamsson
pekkaa@ntnu.no
Norwegian University of Science and Technology, Department of computer and information science, Trondheim, Norway

---

[1] Ilona Kuzmickaja, Xiaofeng Wang, and Daniel Graziotin all contributed equally to the study

# Abstract

It is well acknowledged that innovation is a key success factor in mobile service domain. Having creative ideas is the first critical step in the innovation process. Many studies suggest that customers are a valuable source of creative ideas. However, the literature also shows that adults may be constrained by existing technology frames, which are known to hinder creativity. Instead young children (aged 7-12) are considered digital natives yet are free from existing technology frames. This led us to study them as a potential source for creative mobile service ideas. A set of 41,000 mobile ideas obtained from a research project in 2006 granted us a unique opportunity to study the mobile service ideas from young children. We randomly selected two samples of ideas (N=400 each), one contained the ideas from young children, the other from adults (aged 17-50). These ideas were evaluated by several evaluators using an existing creativity framework. The results show that the mobile service ideas from the young children are significantly more original, transformational, implementable, and relevant than those from the adults. Therefore, this study shows that young children are better sources of novel and quality ideas than adults in the mobile services domain. This study bears significant contributions to the creativity and innovation research. It also indicates a new and valuable source for the companies that seek for creative ideas for innovative products and services.

# 1. Introduction

Creative ideas, as a starting point of any innovation endeavor, play a crucial role for companies who seek competitive advantages in a turbulent marketplace (Cox & Blake, 1991). This is especially true for the companies operating in mobile service sectors and related business. In this paper, *mobile service* is used as an umbrella term to refer to mobile apps, mobile software-as-a-service, hardware and any combinations of them. Several changes occurred in the mobile service market since the Apple *App Store*, the *Google Play* store, and the Nokia *Ovi* store started opening in 2007 (Lane et al., 2010). These platforms have revolutionized the concept of the mobile phone, hosting the release of new content every day. Nowadays, the majority of mobile devices require services such as voice and data services, SMS (Short Message Service), video streaming, location-based services, etc. For the companies in this arena, one key question they need to constantly answer is: *where are the ideas for the next leading mobile services coming from?*

Innovative ideas can come from both inside and outside a company. Companies that utilize external actors and sources in their idea generation processes tend to be more innovative (Laursen & Salter, 2006). Research into the relationship between customer and product innovation maintains that existing customers are often considered a valuable source of creativity and innovation (von Hippel, 1986). The "voice of customer" needs to be heard (Laursen & Salter, 2006). Listening to the voices of customers and observing their behaviors may provide valuable data on unsatisfied needs and point to creative solutions to existing problems. However, for high technology industries such as mobile technology, it has been argued that *ordinary* customers are not a good source for new ideas because "real-world experience of ordinary users is often rendered obsolete by the time a product is developed or during the time of its projected commercial lifetime" (von Hippel, 1986). The widely employed term *lead* users—those users who face needs that will be general in a marketplace several months or years before ordinary users do—is coined in the same study. Von Hippel (1986) argued that lead users are in a better position to provide accurate data on future needs; however, they suffer the same constraints of ordinary customers posed by their real world experience and available technology.

Children, on the other hand, are one group of people who suffer less from the above mentioned constraints (Druin, 2002), and a never-ending source of imagination (Scaife & Rogers, 1998). They are less constrained by existing technology frame due to their little life experience. Still, they somehow remain as a neglected group by market and innovation research, perhaps due to the traditional views of the all-knowing adults and the all-learning children (Druin, 2002). Children are not considered a lead user group as defined by von Hippel (1986), and they cannot even be considered as a customer group due to their lack of purchasing power. However, the studies conducted before the Internet age hint that children, especially young children, are more creative than people from other age groups. Children have been proposed to be part of the processes to design new technologies, in the roles of informants, design partners or even leaders (Druin, 2002, 2010; Read, 2015; Vint, 2005; Yip et al., 2013). The mobile phone industry might really benefit from this neglected source of creative ideas.

*Can young children be a valuable source of creative mobile service ideas*? This is the research question that our study sets out to answer. As far as the authors are aware of, there are no other studies along this line of inquiry. The purpose of this study is to have a better understanding of the voices of young children in terms of creative mobile service ideas. To this end, two sets of mobile service ideas randomly sampled from a larger survey conducted in 2006 were analyzed. One set contained 400 unique ideas expressed by a group of young children of 7 to 12 years old. The other set included 400 distinctive ideas from a group of adults aged from 17 to 50 years. These ideas were analyzed using a conceptual framework of creativity derived from literature, in which creativity is conceptualized as a compound concept with two dimensions: novelty and quality. Both dimensions have two associated constructs: originality and paradigm relatedness for the novelty dimension, and relevance and workability for quality (Dean, Hender, Rodgers, & Santanen, 2006). We tested the difference between the two samples along the above-mentioned dimensions and constructs. The findings of our study empirically demonstrate that young children are actually a valuable source to derive novel ideas that are also of high quality. In contrast, the adults' ideas are deemed to be less novel and of lower quality.

The remaining part of the paper is organized as follows. This section continues with laying out the background and related work. The concept of creativity is investigated in the Conceptualization of Creative Ideas sub-section. The research approach is described in the following section. Then the Result section reports the obtained outcomes of the study, which are further discussed in the light of relevant studies in the Discussion section. The limitations of the study are reflected upon in the same section. The last section concludes the paper and outlines future work.

### 1.1. Mobile Services

According to Alahuhta (2011), mobile services are "*radio communications services between mobile devices while in motion or between such stations and fixed points of services (computer systems/servers)*". The architecture of mobile service systems can be decomposed into three components: 1) Wireless communication infrastructure, 2) mobile terminals and 3) mobile (content) services and apps. The majority of mobile services and apps rely on cellular networks. Although many apps function in offline mode, nowadays users have to be online to benefit from most mobile services.

For end users the most concrete embodiment of mobile services is the mobile device itself. Within a decade a great development has occurred in mobile devices. This development is mainly due to the miniaturization and an increasing level of integration of electronic devices. After the introduction of physical full keyboards to business-oriented mobile phones, the product line has greatly developed in terms of the number of cellular bands, the quality of display, the amount of memory and storage, the variety of data access methods, the capabilities of running mobile applications and the number of features. Desoli & Filippi (2006) presented the evolution of mobile terminals and stated that new upcoming modular devices could satisfy increasing user demand. The size and weight of mobile devices have decreased steadily until the introduction of touch-based smartphones where, due to the new interaction method and improved user experience, the displays have become larger than earlier. The first iPhone sale on June 29$^{th}$, 2007 is a watershed of mobile terminals and the boom of the smartphone era. Additional File 1 is a portrait of the technological evolution of mobile terminals using typical phones of the year as examples.

Mobile content services and apps can be classified in different ways. Alahuhta, Abrahamsson, & Nummiaho (2008) suggested a list of categories from an end-user perspective, including information pull and push, service request, locating persons, objects, identification, etc. The full list of categories and descriptions can be seen in Table 1.

**Table 1. Categorization of mobile services**

| Category | Description |
|---|---|
| Information pull | Retrieving information for some purpose. |
| Information push | Receiving information automatically. |
| Locating (persons/objects) | Locating or following some (nearest) person or object. |
| Communication | Social discussion channel. |
| Service request | Ordering a personal service (possibly based on |

|  | location). |
|---|---|
| Content production | Producing content. |
| Payment | Using mobile device as a means of payment. |
| Identification | Use mobile as an identification device. |
| Other mobile service ideas | Applications that do not fit into other categories. |

Until 2009 the distribution of mobile content services and apps used to be controlled either by device manufacturers or traditional telecom operators. In July 2008 Apple launched the *App Store* to promote mobile applications for the iPhone and iPod mobile handsets. Similarly, Google launched its own store for applications running the Android operating system in October 2008. These are the two most popular and well-known application stores. The application stores attempt to integrate applications closely with devices, so that application downloading is simple and easy and the user experience is optimized (Alahuhta, 2011).

### 1.2. Source of Creative Ideas

The mobile service domain is highly dynamic and innovation plays a crucial role (Siau & Shen, 2003). Obtaining creative ideas is the first step towards innovation. McLean (2005) stated that "*without creative ideas [...] innovation is an engine without any fuel*", and further elaborated that "*no innovation is possible without the creative processes that mark the front end of the process: identifying important problems and opportunities, gathering information, generating new ideas, and exploring the validity of those ideas*" (p.227).

Where do creative ideas come from? Ideas for new products or processes can come from both inside and outside a company. Laursen & Salter (2006) claimed that companies who use external actors and sources in their idea generation processes tend to be more innovative. Customer needs and consumer trends are often a valuable source of innovation. Ideas provided by end users can play a major role in the development of new services. For example, one case study of banking services showed that the customers of the bank proposed ca. 40 new functionalities. In comparison only 7 novel functionalities were proposed by the bank itself (Oliveira & Von Hippel, 2011). Therefore, the "voice of customer" needs to be heard (Laursen & Salter, 2006).

However, for novel products characterized by rapid change such as mobile services, the insights of existing customers and users into new product, process, or service needs and potential solutions are "*constrained by their own real-world experience*" (von Hippel, 1986). As a consequence, customers "*steeped in the present*" are "*unlikely to generate novel product concepts which conflict with the familiar*" (p. 791), as the familiarity with the attributes and uses of existing products affect an individual's ability to conceive novel attributes and uses. It also affects the ability to conceive new product needs, especially in high technology industries (von Hippel, 1986). Von Hippel (1986) coined the term "lead users" of a product whose needs will become general in a marketplace months or years in the future, and who do have real-life experience with novel product or process concepts of interest. They are in a better position than "ordinary" users to provide new product concept and design data.

However, von Hippel (1986) admitted that the insights of lead users could be as constrained to the familiarity as those of other users. Therefore, a natural question to ask is, who are the group of people that are least constrained by their own real-world experience and familiarity with existing technologies? This points our attention to children.

### 1.2.1. Creativity of Children

Several studies investigating creativity based on a person's age have agreed that children are more creative than adults, because they explore the world with "fresh eyes". Instead, rather than producing ideas based on received new information, adults are eliminating information to simplify daily routine (Vint, 2005; von Hippel, 1986, 1988). According to Vint (2005), Land & Jarman (1993) evaluated an individual's creativity over the time. A three-step research has been conducted: 1) in 1968, 1,600 five-year old children were studied and 98% of them were evaluated as creative; 2) in 1973, those children were tested again as they were ten years old – 30% being creative, 3) in 1978, a final test was conducted on the same group of children when they were teenagers (fifteen years old) – only 12% of them were considered creative. On the other hand, the same study evaluated 280,000 adults and only 2% of them were considered creative. Therefore, the younger a person is, the higher the tendency of being creative.

However, how "fresh eyed" are today's children in terms of information technologies? Today's children have been characterized as being *digital natives*, as opposed to their parents and instructors, who are better considered as *digital immigrants* (Prensky, 2001). Recent studies have reported how high is the amount of digital-based experience that children have been exposed. It has been shown that there is an increasing use of computers, Internet, videogames, and mobile devices by children (Subrahmanyam, Greenfield, Kraut, & Gross, 2001). In particular, it has been shown that half of the children in ten U.K. primary schools were already mobile phone owners before 2004 (Davie, Panting, & Charlton, 2004). When smartphone is concerned, a recent research project found that 40% of European boys and 37% of European girls aged 9-12 have a smartphone for private use (Mascheroni & Ólafsson, 2013).

Would the familiarity with information technologies prevent children from being creative, as it puts constraints on adults? Not necessarily. For example, Jackson et al. (2012) investigated children creativity and information technology use, and revealed a certain correlation existing between videogame playing and children creativity. Children have also been proposed to be part of the processes to design new technologies, in the roles of informants and design partners (Scaife & Rogers, 1998; Druin, 2002), and more recent work looked into the potential of children leading the process of design from initial problem formulation to design review and elaboration (Yip et al., 2013). The possibility and effectiveness of children being expert evaluators using heuristic evaluation method has also been explored (Salian, Sim, & Read, 2013). As Read, (2015) argued, the motivation for the involvement of children as active participants and evaluators has been that there is a considerable distance between children and any (adult) expert "guessers", and children act in ways that could not have been predicted by an expert.

In brief, even though there is a seeming tension between children having "fresh eyes" (less constrained by existing technologies) and their being digital natives, evidences in the literature show that the two may not be in conflict but could both boost the creativity of children in the domain of information technologies. Based on the reviewed literature we would expect that, when mobile services are concerned, the ideas coming from young children should be more creative than those generated by adults, therefore young children can be a valuable source of innovation. In this paper we focus on the creativity of young children aged from 7 to 12 years. According to (Piaget, 1953), this is a homogeneous group in terms of their intellectual development. Young children in this age group are able to think abstractly and make rational judgments about concrete, observable phenomena.

### 1.3. Conceptualization of Creative Ideas

Creativity is a multi-faceted, multi-disciplinary concept that is difficult to measure (Piffer, 2012). Over hundred definitions exist for creativity, spanning several disciplines (Hocevar & Bachelor, 1989). According to Sternberg (2006), a component of creativity is imaginative thinking, that is, the ability to see things in novel ways, to recognize patterns and make connections. Nijstad & Paulus (2003) described creativity as "*the development of original ideas that are useful or influential*". Rhodes (1961) suggested that creativity could be an attribute of a process, a product, a person or environmental press, so called four P's model of creativity.

However Dean et al. (2006) argued that, to define idea creativity, it was helpful to differentiate it from the concept of creativity itself. Drawing upon MacCrimmon, Wagner, & Wagner (1994), they defined "*a creative idea as a quality idea that is also novel. That is, it applies to the problem, is an effective and implementable solution, and is also novel*". Based on a literature review of 51 studies on quality, novel and creative ideas, they summarized a conceptual framework of idea creativity as illustrated in Table 2.

**Table 2. Conceptualization of Idea Creativity (adapted from Dean et al. 2006)**

| Dimensions | Constructs |
|---|---|
| Novelty | Originality |
|  | Paradigm Relatedness |
| Quality | Workability |
|  | Relevance |

*Novelty* is considered to be the main dimension of creativity (Dean et al., 2006). A novel idea is rare, unusual, or uncommon (Connolly, Dalgleish, Kalverboer, Hopkins, & Geuze, 1993). According to this definition, the most novel idea is an idea that is totally unique; conversely, the least novel idea is the most common one (MacCrimmon et al., 1994). Dean et al., (2006) warned that when applying the framework, the novelty of any idea must be judged in relation to how uncommon it is in the mind of the idea evaluator or how uncommon it is in the overall population of ideas. *Novelty* can be broken down into two constructs: *originality* and *paradigm relatedness*. Ideas are considered original when they are rare but also have the characteristic of being ingenious, imaginative or surprising. Idea originality ranges from those that are common and mundane to those that are rare and imaginative. *Paradigm relatedness* describes the transformation potential of ideas. It is the degree to which an idea relates to the currently prevailing paradigm, and it is closely related to the concepts of *transformational* and *germinal*.

Based on this understanding, we can formulate the following hypotheses regarding the novelty dimension of idea creativity:

$H_1$: Mobile service ideas expressed by young children are more novel than those by adults.

$H_{1a}$: Mobile service ideas expressed by young children are more original than those by adults.

$H_{1b}$: Mobile service ideas expressed by young children are more transformational in terms of existing paradigm than those by adults.

The second dimension of idea creativity is *quality*, which is further divided into *workability* and *relevance*. An idea is workable (or feasible) if it can be easily implemented and does not violate known constraints. An idea is relevant if it applies to the stated problem and will be effective at solving the problem. There is a third construct suggested by Dean et al. (2006), namely *specificity*. An idea is specific if it is clear and worked out in detail. We excluded the specificity construct in this study, since Dean et al. (2006) recommended that specificity was optional and should be measured only when it is a main focus of a study.

Consequently, we formulate the following hypotheses regarding the quality dimension of idea creativity:

$H_2$: Mobile service ideas expressed by young children are of higher quality than those by adults.

$H_{2a}$: Mobile service ideas expressed by young children are implemented more frequently than those by adults.

$H_{2b}$: Mobile service ideas expressed by young children are more relevant than those by adults.

## 2. Materials and Methods

### 2.1. Sample

The research design has relied on a historical data set. In 2006, a group of researchers from the VTT Technical Research Centre of Finland launched a national research project called *the Idea Movement* (Leikas, 2007). They systematically collected innovative ideas of mobile services from the Finnish citizens through 31 workshops organized in the country in that year. The locations of the workshops included universities, schools, workplaces and even a shopping centre.

The workshops were run in a consistent format, but the targeting participants were grouped by their ages. Therefore the workshops were run separately for children and adults. Each workshop was kicked off with a short introduction to scientific process and idea generation techniques, which was then followed by brainstorming sessions both individually and in groups. Each participant was asked to produce 20 ideas individually and then to form groups of 3-4 people per group. Each group was targeting at generating 100 ideas or more. The researchers acted as the facilitators of the workshops and encouraged the participants not to think about the technology, but to express their concrete needs, and to produce ideas even if they would feel them either "silly" or not realistic in their minds. The ideas were produced in a written format. For this reason, the writing ability of young children did most likely have impact on the extent of how the ideas were expressed. Regardless, young children wrote down their ideas independently without the assistance from the adults present at the workshops.

The ideas were generally 1 to 2 sentences in length, describing a mobile service idea, or expressing a need that one thinks could be fulfilled utilizing mobile technology. Two examples of the ideas are:

"The camera of the phone would have a recognition feature which would recognize persons in photos." (Child idea)

"A composing service which lets one to inspect song patterns that are based on different mathematical forms." (Adult idea)

In total, 41,000 ideas were collected from 2,150 participants, the majority of whom were university students, school children and elderly people. Among the total ideas, 1,800 were from young children aged 7 to 12 years, and 25,300 were from adults aged 17 to 50 years.

In order to decide the meaningful sample sizes for comparing the ideas of young children and adults, we consulted several studies in statistics and organizational research. We set a value of .05 as margin error, as suggested in the guidelines by Bartlett, Kotrlik, & Higgins (2001) and Krejcie & Morgan (1969). We adopted the most conservative value for variance estimation of .50, that is the one usually found for dichotomous variables and will also produce the maximum sample size (Bartlett et al., 2001). Subsequently by employing the Cochran (1977) formula for sample size estimation, we calculated that for a .95 confidence level we needed at least sample size of N=384. On the other hand, the application of a simplified formula from Yamane (1967), which took into account the population size as well, gave us the sample size value N=395. That is, any sample size between 384 and 395 was sufficient for our scope. As a result, we opted to randomly select 400 distinct ideas from the idea set of young children (7-12 years old) and 400 distinct ideas from that of adults (17-50 years old). Our two samples can be freely accessed online[1].

## 2.2. Idea Creativity Assessment

Two independent raters evaluated the creativity of each idea in each sample by employing the framework described in the previous section (as shown in Table 2). Both judges were adult mobile service users and had years of experience with mobile devices. One of the judges had six years experience in mobile service development and was an early adopter of mobile devices. The other rater was a researcher on open innovation and creativity in information systems. Before the rating, the judges were trained on the definition and understanding of the idea creativity concept and measurement scales. The judges were blind to the sources of the ideas, which were ungrouped and randomized.

### 2.2.1. Novelty assessment

The two constructs of novelty, originality and paradigm relatedness, were measured with a Likert scale from 1 to 5. A score of 1 was assigned to least original or least influential. A score of 5 represented the other extremes. The detailed scale descriptions can be seen in Table 3. Following the recommendation of Dean et al. (2006), the novelty score was calculated as the sum of originality and paradigm relatedness, thus ranging in the interval [2,10].

**Table 3. Scales used to evaluate the two constructs of idea novelty**

| Score | Originality level description | Paradigm relatedness level description |
|---|---|---|
| 5 | Surprising, ingenious, not expressed before (rare, unusual) | Paradigm breaking or shifting, introducing several new elements and changing the interactions between mobile service users and mobile technology |

| 4 | Unusual, imaginative | Major paradigm stretching, change the interactions between mobile service users and mobile technology |
| 3 | Interesting, shows some imagination | Moderate paradigm stretching, introducing several new elements |
| 2 | Somehow interesting | Slight paradigm stretching, introducing few new elements |
| 1 | Common, mundane, boring | Paradigm preserving, no influence to future mobile services |

All ideas were generated and collected in 2006. Therefore, the novelty assessment was challenging for the present study. To address this issue, we selected two best-selling mobile phones in 2006[2]: Nokia 1600[3] (130 million sold, see Additional File 1 for its main characteristics) and Nokia 6070[4] (50 million sold). Each rater studied the specifications of each phone to have a clear presentation of what mobile functionalities and services were available at that time. While Nokia had published smartphones already in 2005 (E61) and in 2006 (N95), they were not in a widespread use at the time when the data was collected.

### 2.2.2. Quality assessment

Quality was composed of two constructs: workability and relevance. In terms of workability, the employed data set granted a unique opportunity to go beyond a simple speculation. We could investigate if these ideas were actually implemented after they have been generated in 2006. Any idea that has been developed, produced and marketed in at least some part of the world was considered implemented, therefore workable. On the other hand, ideas that had not been turned into mobile services or apps at all or that were in development but had not been marketed were considered not implemented, therefore not workable. Evaluated in this way, 1/0 dichotomous scores were used. In order to see whether an idea has been implemented or not, information on software, hardware, or service that would match the idea were searched through various online channels. The complete procedure of the workability assessment that we employed is available in Additional File 2.

Relevance was also assessed using the dichotomous scores, to enable the aggregation to the higher quality dimension. A score of 1 meant that an idea was a relevant mobile idea, 0 otherwise. In order to determine if an idea was a relevant mobile service idea, we employed the category list of mobile services (as shown in Table 1). If an idea could be put into the listed categories, it was considered a relevant idea. Otherwise, the idea was considered not relevant. It is worth noting that the ideas that indicate new but un-named types of mobile service were considered relevant and classified under the category "Other mobile service ideas".

The quality score was calculated as the sum of relevance and workability, within the [0,2] score range. It is worth mentioning that, due to the different scales used to measure novelty and quality, we could not aggregate the novelty and quality dimensions as a single creativity score in this study. This is a trade-off we decided to make in order to benefit from the historical nature of the data that allowed us to investigate the actual implementation of these mobile ideas.

## 3. Results

This section provides the analysis outcomes of the two random samples of 400 ideas generated by young children and 400 ideas by adults. It is organized in two sub-sections, Idea Novelty and Idea

Quality. Each sub-section contains the descriptive statistics and the hypothesis testing of each creativity dimension.

## 3.1. Idea Novelty

Additional File 3 contains a list of 3 most novel ideas from the children and adults samples, as well as 3 least novel ideas from the two samples.

### 3.1.1. Descriptive Statistics

Table 4 summarizes the measures of the novelty dimension and its two constructs, originality and paradigm relatedness.

**Table 4. The Novelty of Young Children's and Adults' ideas**

|  | Young Children's Ideas | | | Adults' Ideas | | |
| --- | --- | --- | --- | --- | --- | --- |
|  | Originality | Paradigm relatedness | Novelty | Originality | Paradigm relatedness | Novelty |
| Mean | 3.31 | 3.19 | 6.50 | 2.18 | 2.21 | 4.39 |
| Standard Deviation | 0.75 | 0.75 | 1.26 | 0.81 | 0.75 | 1.45 |
| Median | 3.50 | 3.00 | 6.50 | 2.00 | 2.00 | 4.50 |

Recalling that the score range of the two constructs is [1,5], with respect to both originality and paradigm relatedness, the young children sample presented higher values for the mean and the average rating of the ideas. Additionally, the mean and the median of the adults sample were less than 3, which is the central value of the rating scale. It appeared that the young children on average were able to provide sufficiently original and related ideas, whereas the adults could not.

In Figure 1 and Figure 2, the boxplots of originality and paradigm relatedness show strong differences between the two samples. Regarding originality (Figure 1), the distribution of the averaged scores of the young children's ideas is noticeably greater than that of the adults' ideas. The first quartile of the children sample – i.e., 3 - is greater than the third quartile of the adults sample - i.e., 2.5. Additionally, the children's median value corresponds to the third quartile value of 3.5 and is greater than the adults' median value, by 1.5 units. It should be noted that the values of the outliers in Figure 1 and Figure 2 fall outside 1.5 times the interquartile range and are therefore meaningless in our context. However, it is interesting to notice that 45 ideas from the children sample were rated higher than 4 on average whereas only 2 ideas from the adults sample were rated higher than 4 on average.

For the paradigm relatedness construct in Figure 2, the distribution of the averaged scores of the children's ideas appears to be greater than that of the adults' ideas. The first quartile of the children sample and the third quartile of the adults sample have the same value of 3.5. The median of the children's ideas is greater than that of the adults' ideas by 1 unit.

Recalling the score range [2,10] for the novelty dimension, we see in Table 4 that the children sample outperforms the adults one. The mean and the median values of the children sample, both 6.50, are greater than those of the adults sample, which are 4.39 and 4.50 respectively. As shown in the boxplots of Figure 3, the data distribution seems greater for the children sample. The first quartile of the children sample is equal to the third quartile of the adults sample – i.e., 5.50.

### 3.1.2. Hypotheses Testing

The hypotheses related to novelty ($H_1$, $H_{1a}$, and $H_{1b}$) were implemented as tests for differences between groups. A series of Shapiro-Wilk tests on the samples showed strong evidence for non-normality (p-value < 0.0001 for the two groups, for all the three hypotheses testing). Therefore, the hypotheses were tested with a series of one-tailed Wilcoxon signed rank tests. Table 5 summarizes the tests.

**Table 5. Hypotheses Testing for the Novelty of Children's and Adults' ideas**

| RH | Mean Diff. | Median Diff. | W-value | p-value |
|---|---|---|---|---|
| $H_1$ | 2.11 | 2.00 | 128898 | < 0.0001 |
| $H_{1a}$ | 1.13 | 1.50 | 126255 | < 0.0001 |
| $H_{1b}$ | 0.98 | 1.00 | 122995.5 | < 0.0001 |

At a 0.0001 significance level, there was support to $H_{1a}$ and $H_{1b}$, which state that the mobile service ideas expressed by young children are more original and paradigm changing than those by adults. Finally, at the same significance level, the results also show the evidence for a higher novelty level for the children's ideas with respect to the adults' ideas ($H_1$).

## 3.2. Idea Quality

### 3.2.1. Descriptive Statistics

Table 6 summarizes the measures of the quality dimension and its two constructs, relevance and workability.

**Table 6. The Quality of Children's and Adults' ideas**

|  | Children's Ideas | | | Adults' Ideas | | |
|---|---|---|---|---|---|---|
|  | Relevance | Workability | Quality | Relevance | Workability | Quality |
| Mean | 0.98 | 0.81 | 1.80 | 0.95 | 0.69 | 1.65 |
| Standard Deviation | 0.13 | 0.39 | 0.45 | 0.20 | 0.46 | 0.56 |
| Median | 1.00 | 1.00 | 2.00 | 1.00 | 1.00 | 2.00 |

Recalling that the constructs relevance and workability are binary in the interval [0,1], the two samples provide similar values for the relevance. The mean value was 0.98 for the children sample and 0.96 for the adults one, while the median was 1 for both. Regarding workability, a larger difference between the children sample and the adults one was observed. Although the median was the same for both groups – i.e., 1.00 – the children sample had an average value for workability higher by 0.12 units, whereas the difference in the standard deviations was 0.07. It appeared that the children's ideas on average are more often implemented than the adults' ideas. However, no difference in the relevance of the two groups was noticeable.

A total of 393 over 400 ideas (98%) were evaluated as relevant from the children sample, while 383 over 400 (96%) were considered relevant from the adults sample. Thus, the children provided a slightly higher number of relevant ideas, an increment of 2%.

Out of 400 ideas from the children sample, a total of 324 (81%) have been developed in the past years. On the other hand, from the 400 adults' ideas, 276 (69%) have been implemented. Thus, the children sample presented 12% more workable ideas.

### 3.2.2. Hypotheses Testing

Both relevance and workability constructs followed binomial distribution, because their values represent the number of successes in a sample. Therefore, we tested the one-tailed hypotheses ($H_2$, $H_{2a}$, $H_{2b}$) with a Chi-squared test for proportions of two independent samples. The proportions to be tested were the ones mentioned in the previous sub-section. The hypotheses tests are summarized in Table 7.

**Table 7. Hypotheses Testing for the Quality of Children's and Adults' ideas**

| RH | $X^2$ | .95 Confidence Interval | Sample Estimates | | p-value |
|---|---|---|---|---|---|
| | | | Children | Adults | |
| $H_2$ | 17.5282 | 0.0441, 1.0000 | 0.8975 | 0.8237 | < 0.0001 |
| $H_{2a}$ | 3.4794 | 0.0027, 1.0000 | 0.9825 | 0.9575 | < 0.0500 |
| $H_{2b}$ | 15.4111 | 0.0702, 1.0000 | 0.8125 | 0.6900 | < 0.0001 |

With a p-value less than 0.0001, we found significant evidence for the hypothesis that the children's ideas are more workable than the adults' ideas ($H_{2a}$). With a p-value smaller than 0.05, we found significant support for the hypothesis that the children's ideas are more relevant than the adults' ideas ($H_{2b}$). Finally, with a p-value smaller than 0.0001, we found significant evidence to support the hypothesis that the children's mobile service ideas have higher quality than the adults' ones ($H_2$).

## 4. Discussion

We hypothesized that the mobile service ideas expressed by young children aged 7 to 12 years are more novel ($H_1$) and of higher quality ($H_2$) than those by the adults aged 17 to 50. The results of our study support the two higher-level hypotheses. The general perception suggested by the literature was that the ideas from children are wilder and less realistic therefore less relevant and workable than those of adults. However, our findings empirically demonstrate that this may not be the case for the mobile services domain.

In terms of the two constructs of idea novelty, namely originality and paradigm relatedness, our study supports the two hypotheses $H_{1a}$ and $H_{1b}$, and provided significant evidence that children's ideas are more original and transformational in terms of existing paradigm than those by adults. Therefore, our study supports the creativity literature which claims that children are more creative than adults (Land & Jarman, 1993; Vint, 2005) when creative outcomes are concerned. It also provides further evidence that, even though today's young children are increasingly familiar with and confident in using information technologies, unlike adults, they could still keep their "fresh eyes" and open minds when imagining future digital products and services. The reason behind this could be that young children aged 7 to 12 years are in the intellectual development phase, and their minds are still forming and naturally more open than those of adults (Vint, 2005). However, as it is demonstrated, the "window of opportunity" closes very quickly and they lose their creativity dramatically when they grow up. We should either utilize this source of creativity in time or we need to reflect on and react accordingly to enable children to preserve or adults to "re-learn" creativity.

Regarding the two constructs of idea quality, our results support $H_{2a}$ and found significant evidence that the mobile service ideas expressed by children are implemented more frequently, i.e., they are more workable. In the study of Alahuhta, Abrahamsson, & Nummiaho (2008) it was claimed that 95% of adults' ideas could be implemented. Our study found instead that the percentage of implemented adults' ideas was 69%, which is 26% lower than predicted. We also found significant support to $H_{2b}$, which shows that children's ideas are more relevant than those of adults. The results regarding workability and relevance are surprising in the sense that the intuitive expectation would be on the contrary. One would think the ideas from children are less relevant and workable or at least there shouldn't be significant difference between the two age groups.

As part of the relevance analysis, this study classified the children's ideas using the list of categories from Alahuhta, Abrahamsson, & Nummiaho (2008) (as shown in Table 1), and calculated the percentages of each category (see Additional File 4). We compared the percentages of our analysis of the children's ideas to that of the adults' ideas presented in (Alahuhta et al., 2008). We found that the category with the largest percentage in the children's sample was "Other mobile service ideas" (45.5%), which indicates that the ideas from children are more difficult to be classified under existing categories, therefore, more "out of the box". In comparison, the ideas from adults categorized under this category are 24% of the total adults ideas analyzed in Alahuhta, Abrahamsson, & Nummiaho, (2008).

### 4.1. Limitations of the study

Several limitations, most of which are related to the samples, may threaten the validity of our study. Firstly, there was no detailed demographic data available for the original ideas collected in 2006. We could not know about the children and adults behind these ideas more than their ages and that they are from Finland. This is a threat to generalize our findings to children and adults of other nations. However, Finland is an advanced country in terms of mobile technologies, therefore it can be argued that its population is a good representative of current and potential mobile users. Additionally, we could not know exactly how many ideas were generated by the same persons.

However, since the focus of our study is idea creativity rather than person creativity, the unit analysis is idea, not the person. Therefore, the lack of the demographic data does not present a serious threat, even though the availability of this data would help a better understanding of the generalizability of the results to a larger population.

Another limitation is that in this study we did not include another age group of children from 13 to 16 years old, which is considered teenager group. Although their ideas are also interesting to explore and a comparison could be made between them and the other two age groups, the group was excluded from the investigation. The literature review precisely indicated young children (7-12) as a very creative group yet the most neglected one, mostly due to their lack of purchasing power and the traditional beliefs that relegate adults as "all-knowing" and young children as "all-learning" (Druin, 2002). Meanwhile, the creativity of children has been shown to dramatically decrease when they become teenagers ((Land & Jarman, 1993), as cited by (Vint, 2005)). We were mainly interested in studying the extremely young children as source of creative mobile service ideas. For the same reason, we didn't further divide the adult group into more fine-grained groups. Future studies can build on top of our findings, including the teenager group or focusing on the comparison of children's ideas with those from a sub-group of adults (e.g., from 17 to 25, 25 to 35, etc.).

Finally, due to the research design choice, we could not aggregate novelty and quality scores to a final creativity score for the evaluated ideas. Therefore, our claim that the mobile service ideas expressed by young children (7-12 years old) are more creative than those of adults might be limited. On the other hand, the construct of creativity is expressed in our adopted four sub-hypotheses and measures, which all were empirically proved to be significantly in favor of young children.

# 5. Conclusions

Innovation is a key success factor in any high technology sector, including the mobile service domain. Creative ideas are the first step of innovative products and services. Existing customers, especially lead users, are considered the main sources of creative ideas. However, their creativity is constrained by the available technologies and their life experiences. One potential source of creative ideas, which is somehow overlooked, is young children. This led us to study young children (aged 7-12) as another potential source for creative mobile service ideas. Year 2007 saw the introduction of smartphones, the most significant change to the mobile service domain. A collection of more than 41,000 mobile ideas collected in 2006 granted us a unique opportunity to study the potential of children as the source of creative mobile service ideas. In order to understand if the mobile service ideas expressed by young children are more creative than those by adults, we randomly selected two samples from the idea database of a Finnish research project. One sample was a collection of ideas from young children aged 7 to 12, the other from adults (aged 17-50). We evaluated them using a creativity framework distilled from the literature. The results showed that the mobile service ideas from young children are more novel and have higher quality than those from adults.

The paper offers several interesting findings that would be useful for creativity research as well as mobile service providing companies and designers. The theoretical contribution of our study lies in the six hypotheses, which our study supported significantly. They add to the body of knowledge of creativity study, especially regarding the idea creativity of younger age group, i.e., the "digital natives". The practical implication of our study is that it indicates a new and valuable source for the companies that seek for creative ideas for innovative products and services. Even though the cost of collecting ideas from young children is yet to be better understood, we expect that it should not be more costly than from the adults since the ideas in our samples were collected in the same manner.

Several future studies can be derived from our study. Since in this study the analysis was at the idea level, future studies can analyze the creativity at the person level in terms of mobile service ideas. Future studies can bring in more angles as well, such as the time of implementation, when an idea was implemented after it was uttered, and the popularity of an implemented service, indicated by its download rate.

## Competing Interests

The authors declare that they have no competing interests.

Cox, T. H., & Blake, S. (1991). Managing Cultural Diversity: Implications for organizational competetiveness. *Academy of Management*, *5*(3), 45–56. http://doi.org/10.2307/4165021

Davie, R., Panting, C., & Charlton, T. (2004). Mobile phone ownership and usage among pre-adolescents. *Telematics and Informatics*, *21*(4), 359–373. http://doi.org/10.1016/j.tele.2004.04.001

Dean, D. L., Hender, J. M., Rodgers, T. L., & Santanen, E. L. (2006). Identifying quality, novel, and creative ideas: Constructs and scales for idea evaluation. *Journal of the Association for Information Systems*, *7*(10), 646–698. Retrieved from http://aisel.aisnet.org/jais/vol7/iss10/30

Desoli, G., & Filippi, E. (2006). An outlook on the evolution of mobile terminals: From monolithic to modular multi-radio, multi-application platforms. *IEEE Circuits and Systems Magazine*, *6*(2), 17–29. http://doi.org/10.1109/MCAS.2006.1648987

Druin, A. (2002). The role of children in the design of new technology. *Behaviour & Information Technology*, *21*(1), 1–25. http://doi.org/10.1080/01449290110108659

Druin, A. (2010). Children as codesigners of new technologies: valuing the imagination to transform what is possible. *New Directions for Youth Development*, *2010*(128), 35–43. http://doi.org/10.1002/yd.373

Hocevar, D., & Bachelor, P. (1989). A Taxonomy and Critique of Measurements Used in the Study of Creativity. In J. Glover, R. Ronning, & C. Reynolds (Eds.), *Handbook of Creativity SE - 3* (pp. 53−75). Springer US. http://doi.org/10.1007/978-1-4757-5356-1_3

Jackson, L. A., Witt, E. A., Games, A. I., Fitzgerald, H. E., Von Eye, A., & Zhao, Y. (2012). Information technology use and creativity: Findings from the children and technology project. *Computers in Human Behavior*, *28*(2), 370–376. http://doi.org/10.1016/j.chb.2011.10.006

Krejcie, R. V., & Morgan, D. W. (1969). Determining Sample Size for Research Activities. *Educ Psychol Meas*, *30*(3), 607–610.

Land, G., & Jarman, B. (1993). *Breakpoint and Beyond: Mastering the Future Today*. Leadership 2000 Inc.

Lane, N., Miluzzo, E., Lu, H., Peebles, D., Choudhury, T., & Campbell, A. (2010). A survey of mobile phone sensing. *IEEE Communications Magazine*, *48*(9), 140–150. http://doi.org/10.1109/MCOM.2010.5560598

Laursen, K., & Salter, A. (2006). Open for innovation: The role of openness in explaining innovation performance among U.K. manufacturing firms. *Strategic Management Journal*, *27*(2), 131–150. http://doi.org/10.1002/smj.507

Leikas, J. (2007). Idea movement of aging citizens: lessons-learnt from innovation workshops. In *Universal Access in Human-Computer Interaction (UAHCI)* (pp. 923–931). Beijing, China: Springer Berlin Heidelberg. http://doi.org/10.1007/978-3-540-73283-9_100

MacCrimmon, K. R., Wagner, C., & Wagner, K. R. M. and C. (1994, November). Stimulating Ideas Through Creativity Software. http://doi.org/10.1287/mnsc.40.11.1514

Mascheroni, G., & Ólafsson, K. (2013). *Mobile internet access and use among European children. Initial findings of the Net Children Go Mobile project.* Rome, Italy. Retrieved from http://www.netchildrengomobile.eu/reports/

Figures

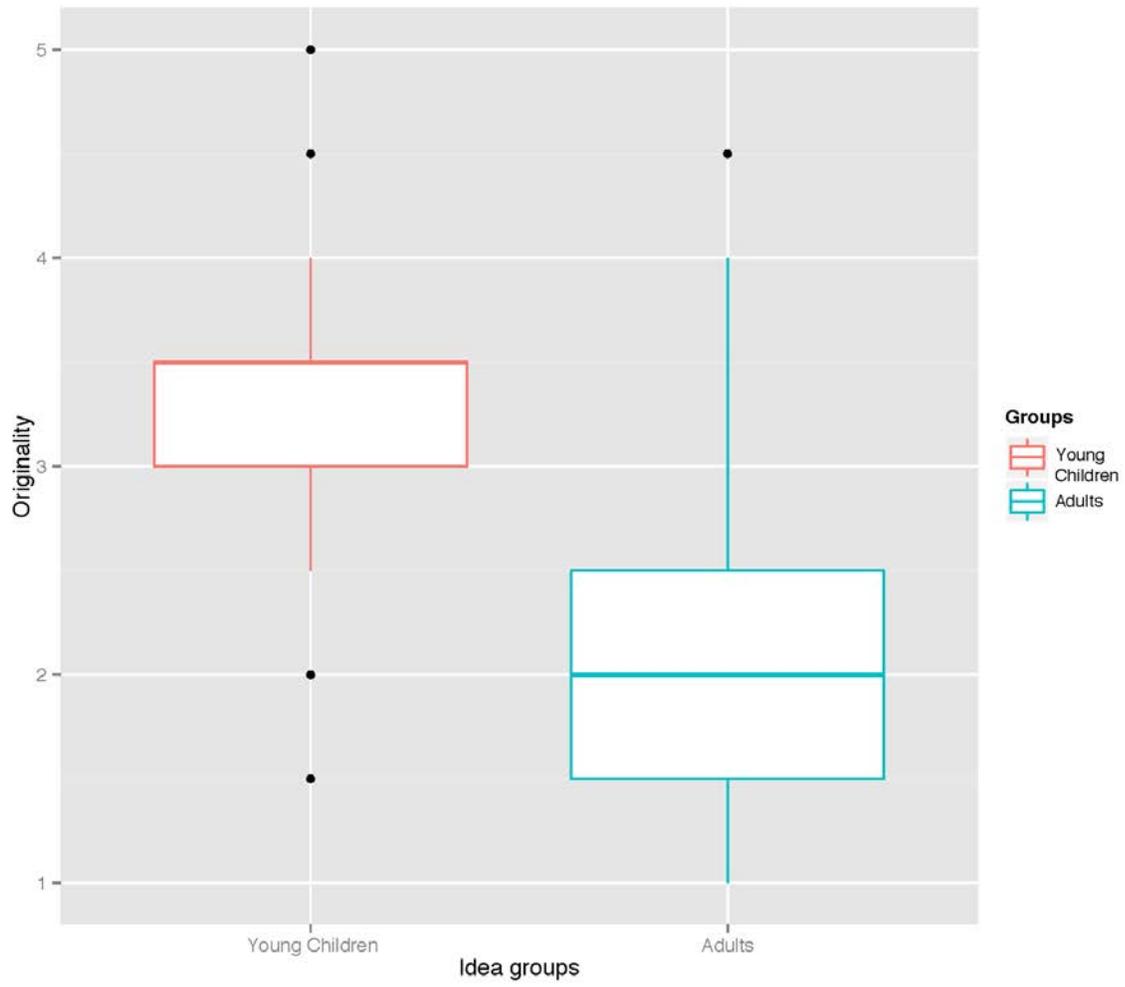

Figure 1. Boxplots for idea originality

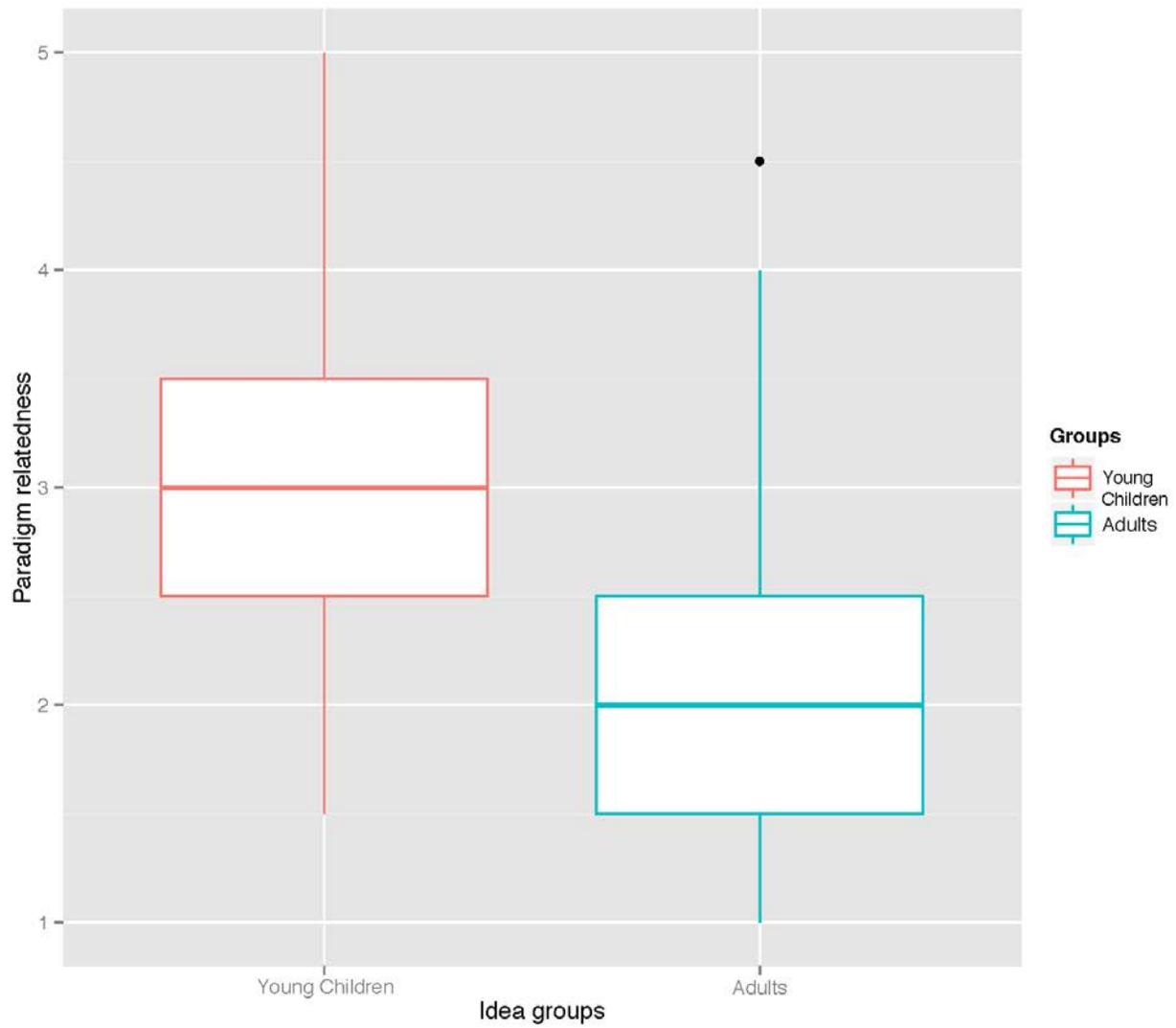

Figure 2. Boxplots for idea paradigm relatedness

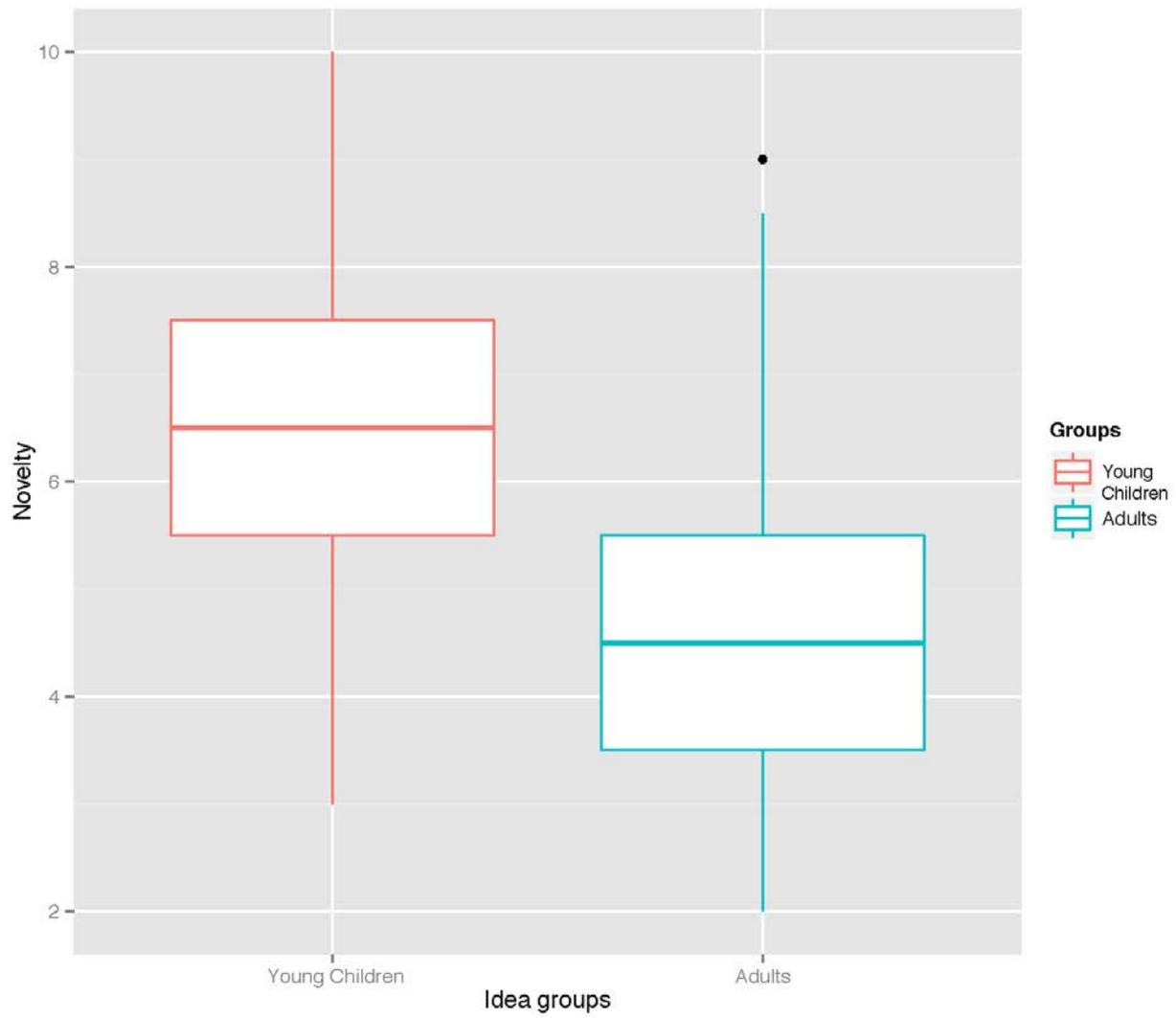

Figure 3. Boxplots for idea novelty

# Additional File 1

Comparison of popular mobile phones in 2006, 2007 and 2012

| Technical characteristics | Nokia 1600 (2006) | Apple iPhone (1st generation, 2007) | Samsung Galaxy SIII (2012) |
|---|---|---|---|
| 2G Network | GSM 900/1800 | GSM 850/900/1800/1900 | GSM 850/900/1800/1900 |
| 3G Network | No | No | HSDPA 850/900/1900/2100 |
| Size (mm, g) | 104x45x17 85g | 115x61x11.6 135g | 136.6x70.6x8.6 133g |
| Input | keyboard | touchscreen | touchscreen |
| Memory | Internal 4MB No Card Clot | Internal 4/8/16 GB No Card Slot | Internal 15/32/64 GB storage, 1 GB RAM microSD |
| Data access | No | GPRS, EDGE, Wi-Fi 802.11b/gm Bluetooth v2.0, USB v2.0 | GPRS, EDGE, HSDPA, Wi-Fi 802.11 a/b/g/n, DLNA, Wi-Fi Direct, Wi-Fi hotspot, Bluetooth v4.0, NFC, |

|  |  |  | USB v2.0 |
|---|---|---|---|
| Camera | No | 2MP, 1600x1200 pixels | 8MP, 3264x2448 pixels, secondary 1.9MP |
| CPU | - | 412 MHz ARM 11 | Quad-core 1.4 GHz Cortex-A9 |
| Battery | Stand-by 450 h Talk time 5h 30min | Stand-by 250h Talk time 8h | Stand-by 590 h (2G)/ 790 (3G) Talk time 21h40min (2G)/ 11h40min (3G) |

## Additional File 2

Procedure to decide the implementation of ideas

**Step 1: Identification of impossible to implement ideas**

The ideas that were obviously impossible to implement, such as "*a cell phone that invalidates gravity*", were directly assigned score 0.

**Step 2: Grouping of ideas**

The remaining ideas were grouped under the categories of pure mobile application, mobile service that needs peripheral device and specific to the features of mobile phones.

**Step 3: Search implementation information**

For pure mobile application ideas we first went on the Apple App Store, the Android Market (now Google Play), and Nokia Ovi Store to see if any application indicated by an idea that might have been developed was there. If the application was not found on any of these platforms, we then went on to try to find it on other types of websites. If this search yielded no results, we would start searching for any type of papers or other publication of possible applications that were still in development, or that were only released in limited regions. This and the other searches were done very similarly through various search engines across the Internet, including *Google*, *Yahoo!*, *Bing*, *Baidu*, *Sogou*, and *Yandex*.

For the ideas that require peripheral devices, we went on the websites of various producers for iPhone peripherals and other brands (such as LG and Samsung), and then resorted on third-party websites (such as *Griffin technology*[2] and *Hammacher*[3]), if nothing was found on the phone manufacturers website. If third-party peripheral developer websites did not give us any results, we would go on and try to find more unconventional places where such an idea might have been developed, this included various forums and personal websites of people that build their own peripherals for cell phones.

---

[2] http://www.griffintechnology.com/

For mobile phone specific ideas, we searched for existing phones and their features to see if they were compatible. If it was not found, we then searched various technology and cell-phone related news websites that might have articles or publications on ideas that were still in development or have been released in a limited quantity.

---

[3] http://www.hammacher.com/

# Additional File 3

Examples of most and least novel ideas

The table below contains a list of 3 most novel and 3 least novel ideas from the samples of the young children and adults. For the ideas of young children, the two raters gave the highest scores (5) to both originality and influence (paradigm breaking) to the three most novel ideas. In the case of adults, instead, none of the ideas was rated with score 5 on both originality and paradigm relatedness by both raters. We selected three ideas of the highest combined novelty scores (their originality scores range from 3 to 5, the paradigm relatedness scores from 4 to 5).

| Source | Most novel ideas | Least novel ideas |
|---|---|---|
| From young children | - "A phone with an AI that one could speak with and ask questions from a selection of voices"<br>- "The phone could speak"<br>- "All locks and security cameras could be checked with the phone" | - "A better calculator for the phone"<br>- "In addition to a wake-up tune, the phone could have a vibration alarm as well"<br>- "The clock of the phone should be more visible" |
| From adults | - "A personal locker that works with your cell phone, no need for a key"<br>- "A remote controlled coffee maker | - "Mobile dictionary"<br>- "Weather report"<br>- "Special day calendar into which |

| | that automatically switches on" | you can mark special days, and the calendar will send you an SMS notification" |
| --- | --- | --- |
| | - "Feature into the cell phone which measures pulse, blood pressure, etc. and starts an alarm if needed" | |

# Additional File 4

Categorization of the young children's ideas and comparison to Alahuhta et al. (2008)

| Category | Description | Young children's ideas (%) | Adults' ideas (%, from Alahuhta et al. 2008) |
|---|---|---|---|
| Other mobile service ideas | Applications that do not fit into other categories. | 45.5 | 24 |
| Service request | Ordering a personal service (possibly based on location). | 17.5 | 4.7 |
| Information push | Receiving information automatically. | 14.75 | 14 |
| Locating (persons/objects) | Locating or following some (nearest) person or object. | 8 | 9.2 |
| Information pull | Retrieving information for some purpose. | 6 | 30 |
| Identification | Use mobile as an identification device. | 4.5 | 3.4 |

| Content production | Producing content. | 2 | 4.3 |
| Payment | Using mobile device as a means of payment. | 1.75 | 4 |
| Communication | Social discussion channel. | 1 | 7 |

Note: The categories are sorted in the order of decreasing percentage per young children's ideas.

---

[1] Accessible here http://dx.doi.org/10.6084/m9.figshare.858906

[2] http://en.wikipedia.org/wiki/List_of_best-selling_mobile_phones#2006

[3] http://www.gsmarena.com/nokia_1600-1188.php

[4] http://www.gsmarena.com/nokia_6070-1433.php